\documentclass[%
reprint,
superscriptaddress,
amsmath,amssymb,
aps,
prl,
floatfix,
]{revtex4-2}

\usepackage{graphicx}
\usepackage{dcolumn}
\usepackage{bm}
\usepackage{physics}
\usepackage{soul}
\usepackage{times}
\usepackage[breaklinks,colorlinks,urlcolor=blue,citecolor=blue,linkcolor=blue]{hyperref}
\usepackage[mathlines]{lineno}

\usepackage{amsfonts,amssymb,amsmath,mathbbol}              
\usepackage{graphics,epsfig,ulem}
\usepackage{comment}

\begin{document}

\title{Protected Gapless Edge States In Trivial Topology} 
 
\author{Yun-Chung Chen }
\affiliation{Department of Physics and Center for Theoretical Physics, National Taiwan University, Taipei, Taiwan 10607}

\author{Yu-Ping Lin}
\affiliation{Department of Physics, University of Colorado, Boulder, Colorado 80309, USA}
 
\author{Ying-Jer Kao}
\email{yjkao@phys.ntu.edu.tw}
\affiliation{Department of Physics and Center for Theoretical Physics, National Taiwan University, Taipei, Taiwan 10607}
\affiliation{Physics Division, National Center for Theoretical Science, Taipei, Taiwan 10607}

\begin{abstract}              
Bulk-boundary correspondence serves as an important feature of the strong topological insulators, including Chern insulators and $Z_2$ topological insulators. Under nontrivial band topology, the protected gapless edge states correspond to the Wannier obstruction or Wilson-loop winding in the bulk. Recent studies show that the bulk topological features may not imply the existence of protected gapless edge states.
Here we address the opposite question: Does the existence of protected gapless edge states necessarily imply the Wannier obstruction or Wilson-loop winding?
We provide an example where the protected gapless edge states arise without the aforementioned bulk topological features. This \textit{trivialized topological insulator}  belongs to a new class of systems with non-delta-like Wannier functions. Interestingly, the gapless edge states are not protected by the crystalline symmetry; instead the protection originates from the mirror antisymmetry, a combination of chiral and mirror symmetries. Although the protected gapless edge states cannot be captured by the bulk topological features,  they can be characterized by the spectral flow in the entanglement spectrum.

\end{abstract}

\maketitle

\vspace{-2ex} 

\textit{Introduction.---}Topological insulator (TI) is a class of materials that exhibits nontrivial boundary phenomena \cite{Kane2005,Hasan2010,Qi2008,Qi2011,Xu2006,Fu2007,Fu2007_2}.
The prototypical examples are the strong TIs with gapless surface states, which are protected by the internal symmetry, associated with each boundary, and can be classified by $K$ theory \cite{Schnyder2008,Kitaev2009,Ryu2010}. 
The protecting symmetry can  be generalized to crystalline symmetries to define topological crystalline insulators (TCIs) \cite{Teo2008,Fu2011,Hsieh2012,Liu2014,Po2017,Neupert2018}. 
One example in this class is the mirror Chern insulator in three dimensions, which has gapless surface states under the protection of mirror symmetry \cite{Teo2008,Hsieh2012,Liu2014}. 
The gapless states can only arise at mirror-invariant surfaces, while all other surfaces are gapped. 
This type of bulk-boundary correspondence (BBC) is very different from that of strong TIs, where all the surface states are gapless.
Recently, a new class of TCIs, the higher-order topological insulators (HOTIs), has been identified~\cite{Neupert2018,Song2017,Benalcazar2017,Benalcazar2017_2,Langbehn2017,Schindler2018}, where the protected boundary states live on the lower-dimensional surfaces.
HOTIs have been now observed in many models and realized in solid-state, photonic and phononic systems \cite{Song2017,Benalcazar2017,Langbehn2017,Schindler2018,Khalaf2018,vanMiert2018,Geier2018,Trifunovic2019,Trifunovic2020,Ezawa2018,Ren2020,Chen2020,Park2019,Schindler2018_2,Xie2018,Ni2019,Chen2019}. 
Another interesting example is the class of systems with fractional corner charges (FCCs) \cite{Benalcazar2017,Benalcazar2017_2,Benalcazar2019,Schindler2019,Watanabe2020,Naito2022,Takahashi2021,Watanabe2021,Lin2021}. 
This type of systems does not have midgap corner states in general, although the total charge accumulations at the corners are quantized.

One may wonder whether there is a direct BBC between the bulk topology and surface responses in TCIs. 
The concept of topology has to be defined first in general systems.
A widely-accepted definition is the Wannierizability of the ground state many-body wave function, including elementary band representations (EBRs) \cite{Bradlyn2017,Cano2018,Cano2018_2} and symmetry indicators \cite{Po2017,Po2020,Khalaf2018_2}. 
A topological insulator cannot be represented by the exponentially-localized symmetric Wannier functions of the crystalline symmetry group \cite{Bradlyn2017,Cano2018}. 
This type of definition is physically clear: a topological insulator is different from an atomic  insulator defined in the atomic limit. 
On the other hand,  $K$ theory offers a generally different formalism from this classification. A prototypical example of this difference occurs in the Su-Schrieffer-Hegger (SSH) model. While both phases of the model are Wannierizable~\cite{Su1980}, a mismatch of the atomic and Wannier centers turns a phase into an obstructed atomic insulator (OAI) \cite{Bradlyn2017,Benalcazar2017,Benalcazar2019}. Despite the trivial topology, OAIs can often exhibit FCCs \cite{Benalcazar2017,Benalcazar2019} and are relatively topological in  $K$ theory. Interestingly, one can also find the Wannierizable topological systems that are considered relatively trivial in  $K$ theory. These systems are known as the fragile topological systems \cite{Po2018,Bradlyn2019,Song2020,Liu2019,Bouhon2019,Hwang2019,Po2019,Song2019,Wang2019,Kobayashi2021,Kooi2019,Wieder2018,Alexandradinata2020,Song2020_2}, as opposed to the stable topological systems with robust bulk-boundary correspondence. The trivialization of these systems can be achieved by adding trivial bands into the occupied space, thereby turning the whole occupied space Wannierizable \cite{Po2018}.

It is now clear that there is a difference between the concepts of Wannierizability and physical responses. 
For the stable TCIs with protected gapless surface states, one always encounters the Wannier obstruction in the construction of exponentially-localized symmetric Wannier functions~\cite{Khalaf2018_2,Song2018}. 
In contrast,  TCIs which have no protected gapless surface states under Wannier obstruction are fragile topological insulators~\cite{Liu2019}.
Despite the absence of the stable protection,  gapless surface states can still appear in the fragile topological insulators and lead to nontrivial responses \cite{Fu2011,Alexandradinata2014_2,Alexandradinata2016_2,Wang2019,Alexandradinata2020,Kobayashi2021}. Nevertheless, the fragile topological systems offer the examples where the bulk topology does not necessarily lead to protected gapless surface states. An interesting question thus arises in the opposite direction: Is it also possible to find  systems hosting protected gapless surface states, yet with trivial topology? 

\begin{figure}
\begin{center}
{\includegraphics[width=8.6cm]{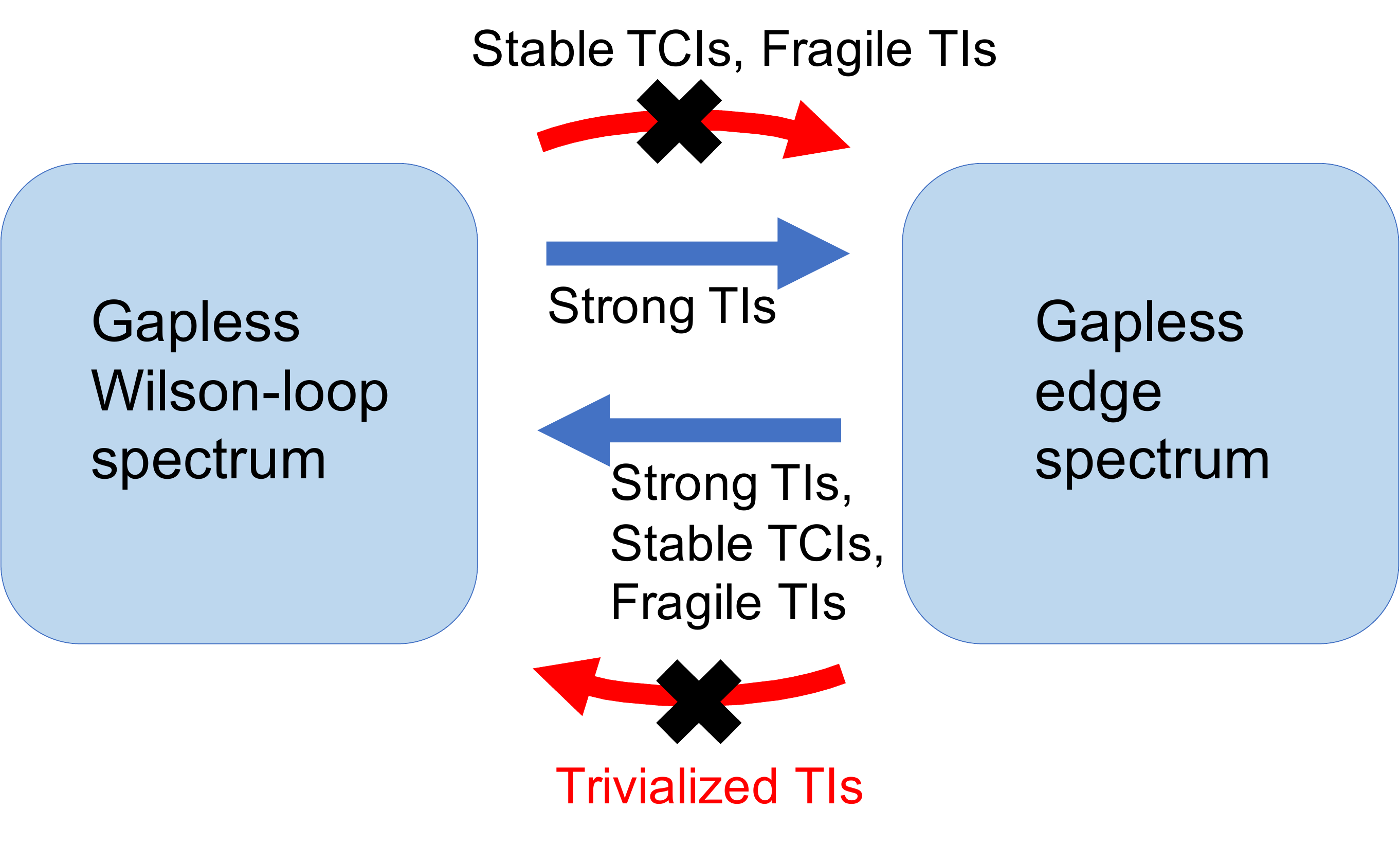}}
\caption[]{Different types of topological insulators under the bulk-boundary correspondence. The arrow corresponds to the direction of the correspondence.}
\label{fig1}
\end{center}
\end{figure}

In this Letter, we address this question by explicitly constructing an example of such systems, termed the trivialized topological insulator (TTI).
This system host protected gapless edge states, while the Wilson-loop winding and Wannier obstruction are absent.
The gapless edge states in the TTI are protected by the combination of the mirror ($\mathcal{M}$) and the chiral  ($\mathcal{S}$) symmetries, or the \textit{mirror antisymmetry} ($\tilde{\mathcal{M}}$). 
Interestingly, TTIs also serve as  counterexamples to the correspondence between Wilson-loop and edge energy spectra \cite{Fidkowski2011,Neupert2018,Benalcazar2017,Taherinejad2014,Taherinejad2015,Varnava2020,Liu2014_2,Alexandradinata2016,Olsen2020}. 
The protected gapless edge states appear even though the Wilson-loop spectrum is gapped. 
We summarize the comparison of TTI and  other types of topological systems in Fig.~\ref{fig1}.
Strong TIs, such as Chern insulators and $Z_2$ topological insulators, necessarily obey the most stringent bulk-boundary correspondence.
On the other hand, it is possible to find the gapped surface states in stable TCIs and fragile TIs although the Wilson-loop spectrum has protected winding.
Due to the Wannierizability, the TTI cannot be captured by the Wilson-loop spectrum and the theory of EBRs. 
Nevertheless, the Wannier functions of the TTI cannot be adiabatically deformed into  delta-functions localized on the atoms. 
The TTI thus belongs to a new class of non-delta-function insulators \cite{Nelson2021,Nelson2021_2,Schindler2021}. 
For these non-delta-function insulators, the conditionally robust surface states exist under the delicate topology \cite{Nelson2021_2}. In contrast, the TTI hosts stable protected gapless edge states, thereby presenting more promising  applications in real materials.

\textit{Model.---} Our point of departure is the 2D time-reversal ($\mathcal{T}$)-invariant $Z_2$ topological insulators \cite{Kane2005,Bernevig2006}.
While this model is chosen specifically to exemplify our analysis, it should be noted that similar results can also be produced in  other models.
Consider a four-band tight-binding model on the square lattice
\begin{eqnarray} \label{equation 1}
H=&&-t\sum_{\langle i,j\rangle,a,b}\nu_{ab}c^{\dagger}_{i,a}c_{j,b}+i\lambda_\textrm{NSO}\sum_{\langle i,j\rangle,a,b}\mu_{ab}c^{\dagger}_{i,a}\bm{s}\cdot\bm{d}_{ij}c_{j,b}\nonumber\\
&&+m\sum_{i,a}\xi_{a}c^{\dagger}_{i,a}c_{i,a}.
\end{eqnarray}
Here $c^{\dagger}_{i,a}=(c^{\dagger}_{i,a,\uparrow},c^{\dagger}_{i,a,\downarrow})^T$ is the creation operator of a spinful electron at site $i$. The orbitals $a=A$ or $B$ have the same irreducible representation $\bar{E}_1$ at the Wyckoff position $a$ of the double space group $P4mm$ \cite{Bradlyn2017,Bilbao}. 
The first term is the nearest-neighbor hopping with the strength $t>0$ and the orbital Pauli matrix $\nu_{ab}=\sigma^z_{ab}$.
The second term is the nearest-neighbor spin-orbit coupling, which couples the orbital $\mu_{ab}=\sigma^x_{ab}$ and spin $\bm{s}=(s_x,s_y)$ Pauli matrices with the unit displacement vector $\bm{d}_{ij}=\bm{d}_j-\bm{d}_i$ from site $i$ to site $j$.
The last term is the on-site energy with $\xi_a=\pm1$ for orbitals $a=A$ and $B$. 
For $0<m<4t$ and $-4t<m<0$, this model shares the same features as the BHZ model with the nontrivial $Z_2$ index \cite{Bernevig2006}.
We analyze this model since the spin-orbit coupling term is symmetric with respect to the lattice. 
The parameters are set as $t=-\lambda_\textrm{NSO}=0.5$ throughout our analysis.
All the numerical results are performed using the P\textsubscript{YTH}TB package \cite{pythTB}.

\textit{Edge theory.---} It is well-known that the $Z_2$ topological insulators have $\mathcal{T}$-protected helical gapless edge states \cite{Wu2006}.
Meanwhile, the gapless edge states of  stable TCIs are protected by the crystalline symmetry. 
Here we show that the combination of a chiral and a crystalline symmetry can also protect the gapless edge states.
Furthermore, the positions of the edge Dirac points are strictly protected at the high-symmetry points. 

Consider a general 2D system with protected gapless edge states at the boundaries. For example, the $Z_2$ topological insulator \eqref{equation 1} has protected gapless edge states due to the nontrivial $Z_2$ index. 
Fourier transforming the lattice model and expanding around a high-symmetry point, we obtain a low-energy effective Dirac Hamiltonian~\cite{Khalaf2018,Geier2018,Hwang2019,Khalaf2018_2}
\begin{eqnarray} \label{equation 2}
\mathcal{H}=\bm{k}\cdot\bm{\Gamma}+\lambda M
\end{eqnarray}
at the relative momentum $\bm{k}$. Here the three matrices $\bm{\Gamma}=(\Gamma_x,\Gamma_y)$ and $M$ anticommute with each other and square to identity $I$, where $M$ corresponds to the bulk mass term in the energy $E(\bm{k})=\pm\sqrt{\bm{k}^2+\lambda^2}$.
The minimal Dirac Hamiltonian describes the low-energy physics of the  Altland-Zirnbauer tenfold symmetry  classes~\cite{Khalaf2018}. 
The edge theory can be obtained via a projection~\cite{Khalaf2018,Geier2018,Trifunovic2019,Hwang2019,Khalaf2018_2}
\begin{eqnarray} \label{equation 3}
h(\bm{k})=P_+(\bm{n})\mathcal{H}P_+(\bm{n}),
\end{eqnarray}
where $P_+(\bm{n})=\frac{1}{2}(1-i\bm{n}\cdot\bm{\Gamma}M)$ is the projector with the unit normal vector $\bm{n}$ of the edge. This yields a gapless Hamiltonian $h(\bm{k})=k_{\parallel}\gamma(\bm{n})$ with $\gamma(\bm{n})=P_+(\bm{n})\bm{\Gamma}P_+(\bm{n})$ at an edge high-symmetry point. Note that the edge Dirac points at all the high-symmetry points are degenerate.
Importantly, the matrix
\begin{eqnarray} \label{equation 4}
\tilde{\mathcal{M}}_{\bm{z}\times\bm{n}}\equiv i\bm{n}\cdot\bm{\Gamma}M
\end{eqnarray}
with the out-of-plane unit vector $\bm{z}$ serves as a representation of the mirror antisymmetry
\begin{eqnarray} \label{equation 5}
\tilde{\mathcal{M}}_{\bm{z}\times\bm{n}}\mathcal{H}(k_{\perp},k_{\parallel})\tilde{\mathcal{M}}_{\bm{z}\times\bm{n}}^{-1}=-\mathcal{H}(k_{\perp},-k_{\parallel}).
\end{eqnarray}
Here $k_{\perp}$ and $k_{\parallel}$ are the momenta normal and parallel to a properly chosen edge, respectively. A crystalline antisymmetry is actually a combination of the chiral and crystalline symmetries \cite{Geier2018,Trifunovic2019,Shiozaki2014}. In our case, the mirror antisymmetry $\tilde{\mathcal{M}}_{\bm{z}\times\bm{n}}$ combines the chiral symmetry and the mirror symmetry which leaves the edge invariant. 
Since $P_+(\bm{n})\Pi P_+(\bm{n})=0$ for any matrix $\Pi$ satisfying $\{\Pi,\tilde{\mathcal{M}}_{\bm{z}\times\bm{n}}\}=0$, any mirror-antisymmetric perturbation $\Delta H$ leaves the edge theory gapless.
This result is consistent with the BBC obtained from  $K$ theory \cite{Geier2018,Trifunovic2019,Hsieh2014}. 

We note that the 2D mirror-symmetry protected TCI in the Altland-Zirnbauer class AIII belongs to a larger set of systems with mirror antisymmetry. While the gapless edge sates are protected by the mirror antisymmetry, the protection is not supported by the mirror symmetry alone once the chiral symmetry is broken \cite{Geier2018,Trifunovic2019,Hsieh2014}.
Since the midgap corner states can be related to a specific edge with gapless edge states~\cite{Geier2018}, the second-order topology cannot be protected by the mirror symmetry alone, either \cite{Suppl}. In fact, the system is topologically trivial if it only preserves the mirror symmetry.

Interestingly, our model becomes a fragile TI when the inversion ($\mathcal{I}$) symmetry is preserved alone~\cite{Hwang2019,Wieder2018}. 
The Wannier obstruction of the fragile topology can be characterized by the relative winding of the Wilson-loop spectrum, which is protected by the $\mathcal{I}$ symmetry \cite{Hwang2019,Alexandradinata2014}.
However, by adding appropriate atomic bands, the occupied bands become Wannier representable \cite{Hwang2019,Wieder2018}. 
Since the fragile topology can not protect the stable gapless edge and corner states, neither can the $\mathcal{I}$ symmetry. 
Therefore, the second-order topology cannot be protected by the mirror or inversion symmetry alone, contrary to the claim in the  literature \cite{Ezawa2018,Ren2020,Chen2020}.
Instead, the midgap corner states are protected essentially by the mirror antisymmetry.

\textit{Gapless edge states in trivial topology.---}
%
We now address the central question of this work: Does the existence of protected gapless edge states imply the Wannier obstruction or Wilson-loop winding? The positive answer is tempting according to a proposed correspondence between the Wilson-loop and edge energy spectra~\cite{Fidkowski2011,Neupert2018,Benalcazar2017,Taherinejad2014,Taherinejad2015,Varnava2020,Liu2014_2,Alexandradinata2016,Olsen2020}. However, we show by explicit construction that there exist counterexamples which turn the answer negative.

The correspondence theorem states a correspondence between the winding of Wilson-loop spectrum and the protected gapless edge states~\cite{Fidkowski2011}.
The hybrid Wannier center, corresponding to the Wilson-loop spectrum, is related to the edge spectrum through a continuous transformation.
However, the continuous transformation can not protect the gapless points in general, even if it respects the symmetry~\cite{Alexandradinata2016}. 
For the strong TIs, the theorem is satisfied trivially since both the Wilson-loop and edge spectra are gapless.
On the other hand, one can find gapped surfaces in some stable TCIs while the Wilson-loop spectra remain gapless, such as in the axion insulators \cite{Varnava2020,Wieder2018}. This property is also shared by the fragile topological systems.
Therefore, the correspondence from the gapless Wilson-loop spectrum to the gapless edge spectrum can break down completely.

\begin{figure}
\begin{center}
\includegraphics[width=\columnwidth]{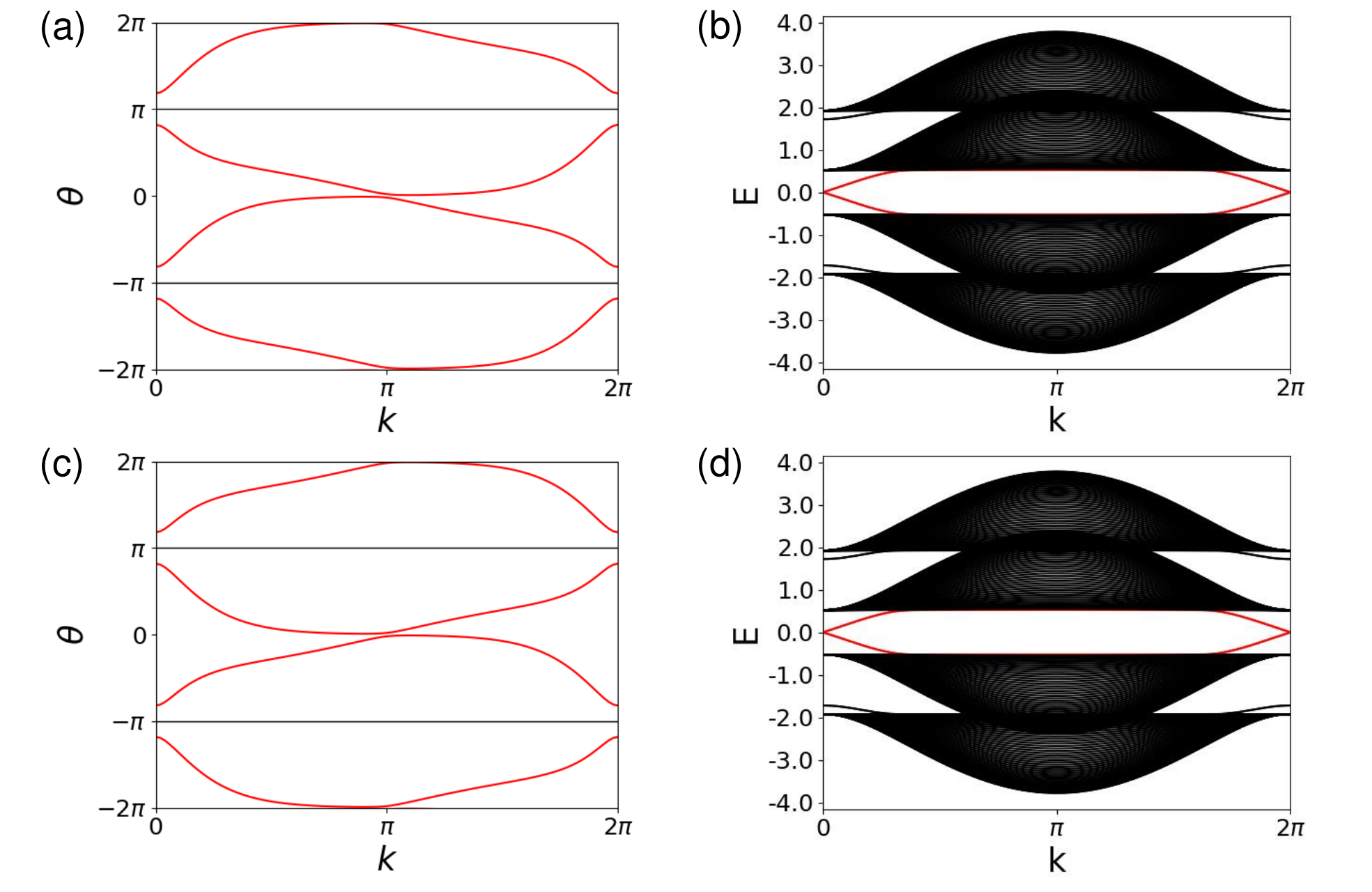}
\caption[]{Protected gapless edge states in the TTI. Here we consider the model (\ref{equation 6}) with $m=1.0$ and $B_z=-\lambda_{\textrm{SO}}=0.7$. (a) and (c) show the Wilson-loop spectra in the $x$ and $y$-direction, respectively. (b) and (d) show the protected gapless edge states (red) from nanoribbon energy spectra in the $y$ and $x$-direction, respectively.}
\label{fig2}
\end{center}
\end{figure}

One may wonder if the correspondence still holds in the opposite direction, as claimed in \cite{Varnava2020}. 
The TTI, where the mirror antisymmetry protects the gapless edge states while the Wilson loop is unwinded, provides an counterexample to this claim.
To construct the TTI, we add two extra terms to the $Z_2$ topological insulator (\ref{equation 1}) to break the protecting symmetries
\begin{eqnarray} \label{equation 6}
H_\textrm{TTI}=&&H+\lambda_\textrm{SO}\sum_{i,a,b}\zeta_{ab}c^{\dagger}_{i,a}s_zc_{i,b}\nonumber\\
&&+B_z\sum_{i,a}c^{\dagger}_{i,a}s_zc_{i,a}.
\end{eqnarray}
The first term is an on-site spin-orbit (SO) coupling with $\zeta_{ab}=\sigma^y_{ab}$. 
This term breaks the $\mathcal{I}$ and $C_2\mathcal{I}$ symmetries and trivializes the mirror Chern number from the in-plane mirror symmetry. 
The second term applies an out-of-plane magnetic field, which breaks the $\mathcal{T}$ symmetry and trivializes the $Z_2$ index.
The fragile topology is also broken due to the loss of protecting $\mathcal{I}$ or $C_2\mathcal{T}$ symmetries. 
Since the mirror antisymmetries $\tilde{\mathcal{M}}_x$ and $\tilde{\mathcal{M}}_y$ are preserved, the existence of protected gapless edge states is guaranteed. 
Figure~\ref{fig2} shows the Wilson-loop spectrum and the nanoribbon energy spectrum for $m=1.0$ and $B_z=-\lambda_{\textrm{SO}}=0.7$. 
The gapless edge states are protected by the mirror antisymmetries $\tilde{\mathcal{M}}_x$ and $\tilde{\mathcal{M}}_y$, but the Wilson-loop winding is no longer protected. 

\begin{figure}
\begin{center}
{\includegraphics[width=8.6cm]{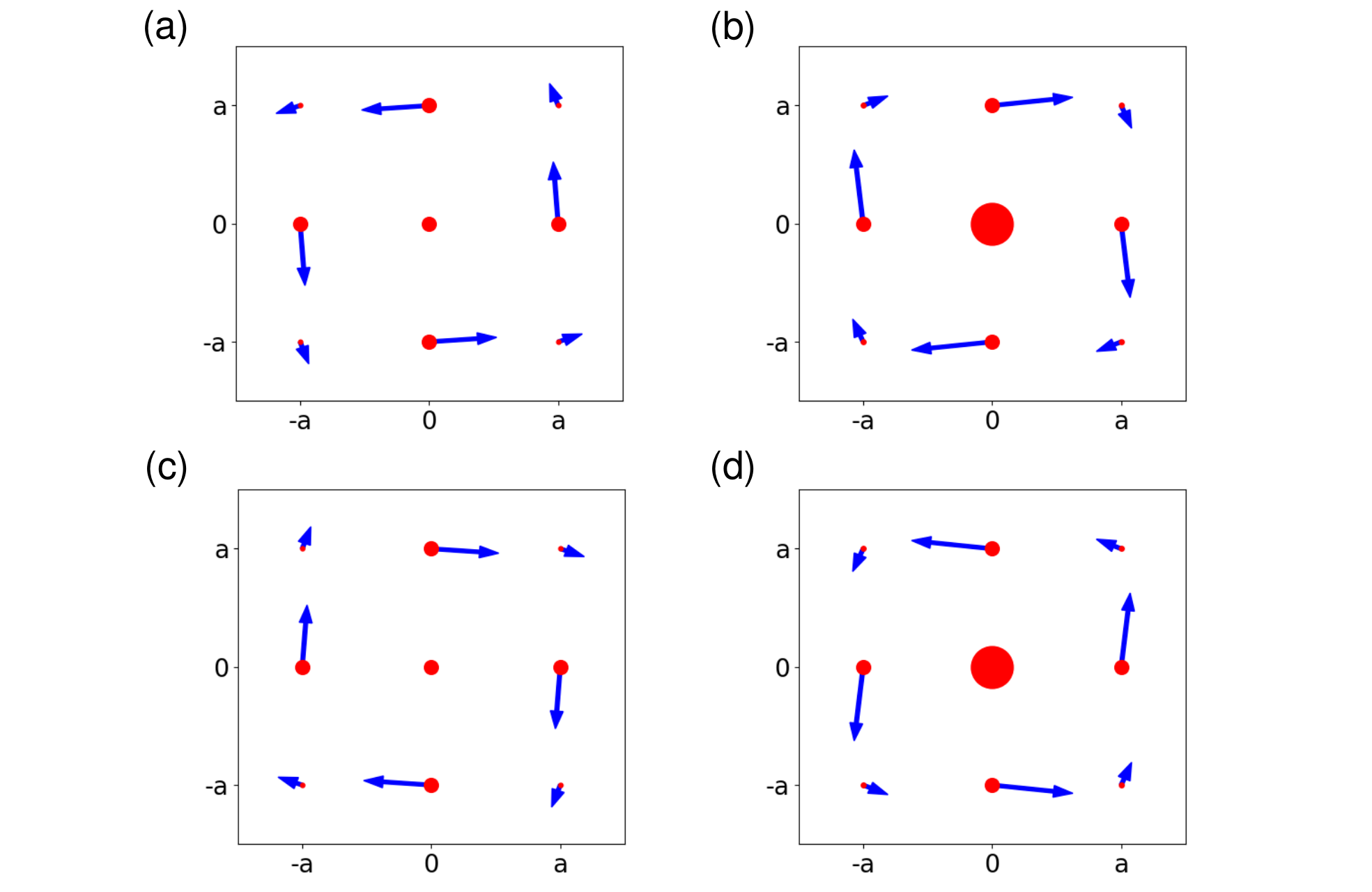}}
\caption[]{Wannier functions of the TTI around the Wanner centers. Each plot corresponds to a single Wannier function. Here we  plot only the nine neighboring atoms, although the results are more extended. The size of the red spot is proportional to the probability of the Wannier function. The blue arrows represent the magnitude and direction of the spins in the $x$-$y$ plane. (a) and (b) illustrate one of the valence band Wannier functions $|W_{\bm{0},1}^v\rangle$ projected to orbital $A$ and $B$, respectively. (c) and (d) are the corresponding conduction band Wannier function $|W_{\bm{0},1}^c\rangle$ projected to orbital $B$ and $A$, respectively. The Wannier functions in (a) and (c), (b) and (d) form mirror pairs.}
\label{fig3}
\end{center}
\end{figure}

Although the $C_4$ symmetry is preserved, it does not protect the topology and imposes no obstruction to constructing the $C_4$-symmetric Wannier functions. 
Having studied the symmetry representations and explicitly constructed the exponentially-localized Wannier functions, we find that the Wannier functions are localized at the Wyckoff position $a$ with irreducible representations ${}^{1}\bar{E}_1$ and ${}^{2}\bar{E}_1$ ( Fig.~\ref{fig3})~\cite{Suppl}. The Wannierizability clearly confirms the trivial topology in the TTI.
Note that our Wannier functions satisfy the mirror antisymmetry
\begin{eqnarray} \label{equation 7}
|W^c_{\bm{0},n}\rangle=A_{\tilde{\mathcal{M}}}|W^v_{\bm{0},n}\rangle.
\end{eqnarray}
Here $|W^c_{\bm{0},n}\rangle$ and $|W^v_{\bm{0},n}\rangle$ are the conduction and valence band Wannier functions at $\bm{R}=\bm{0}$ with band index $n$, and $A_{\tilde{\mathcal{M}}}$ is the mirror antisymmetry operator in real space.

Based on our analysis, we conclude that the TTI hosts protected gapless edge states while remaining topologically trivial. This serves as a counterexample to the correspondence theorem~\cite{Fidkowski2011,Neupert2018,Benalcazar2017,Taherinejad2014,Taherinejad2015,Varnava2020,Liu2014_2,Alexandradinata2016,Olsen2020}. The essence lies in the role of mirror antisymmetry, which protects the gapless edge states but leaves the Wannier obstruction and Wilson-loop winding unprotected. Therefore, the search for TTIs may be focused on the class of mirror-antisymmetric insulators.

\textit{Multicellularity.---}We have learned that the TTI is Wannierizable and topologically trivial. However, the protected gapless edge states make it inequivalent from the trivial atomic insulators. The inequivalence can be understood from the perspective of the Wannier functions. In the trivial atomic insulators, the Wannier functions can be localized to the delta functions \cite{Po2017,Bradlyn2017}. However, the delta-function localization can never be achieved in the TTI. Due to this "multicellularity", the TTI belongs to a new class of non-delta-like insulators \cite{Nelson2021,Nelson2021_2,Schindler2021}, which can not be adiabatically connected to the trivial atomic insulators.

The origin of multicellularity is again attributed to the mirror antisymmetry. In the TTI, the Wannier functions are mirror-antisymmetric according to Eq.~(\ref{equation 7}). An antisymmetric pairing condition
\begin{eqnarray} \label{equation 8}
\langle W^v_{\bm{R},m}|A_{\tilde{\mathcal{M}}}|W^v_{\bm{0},n}\rangle=0
\end{eqnarray}
can be imposed on the Wannier functions for any valence band indices $n,m$ and positions $\bm{R}$~\cite{Suppl}. Here we also assume the half-filling condition. Since the Hilbert space is highly constrained under this condition, the multicellular Wannier functions can not be adiabatically connected to the delta-function ones in the trivial atomic insulators. An inequivalence is thus imposed by the mirror antisymmetry between the TTI and trivial atomic insulators.

The adiabatic inconnectivity can be demonstrated by considering the corresponding Bloch states. For the Wannier functions which satisfy the antisymmetric pairing condition, the corresponding valence band Bloch states $|u_n(\bm{k})\rangle$ obey the mirror antisymmetry
\begin{eqnarray} \label{equation 9}
\tilde{\mathcal{M}}P(R\bm{k})\tilde{\mathcal{M}}^{-1}=I-P(\bm{k}).
\end{eqnarray}
Here $P(\bm{k})=\sum_n|u_n(\bm{k})\rangle\langle u_n(\bm{k})|$ is the projector onto the valence bands and $R$ represents the  mirror  reflection.
The Bloch states further correspond to a flat band Hamiltonian
\begin{eqnarray} \label{equation 10}
H_\textrm{flat}(\bm{k})=I-2P(\bm{k}),
\end{eqnarray}
which is invariant under the mirror antisymmetry
\begin{eqnarray} \label{equation 11}
\tilde{\mathcal{M}}H_\textrm{flat}(R\bm{k})\tilde{\mathcal{M}}=-H_\textrm{flat}(\bm{k}).
\end{eqnarray}
This establishes the correspondence between the set of Wannier functions and the flat band Hamiltonian. The topology of the Wannier functions can then be proved by contradiction. If one can continuously deform the exponentially-localized Wannier functions into the delta functions, the corresponding flat band Hamiltonian can also be adiabatically deformed into the trivial one without crossing a band gap. Due to the presence of protected gapless edge states, this is impossible for the TTI under the mirror antisymmetry.

The TTI shares similar properties to  other non-delta-like insulators, such as those with returning Thouless pump and noncompact atomic insulators \cite{Nelson2021,Nelson2021_2,Schindler2021}. However, there exist fundamental differences in between. First, the TTI has stable topology instead of delicate topology \cite{Nelson2021,Nelson2021_2}. The protected gapless edge states are stable under the addition of atomic bands, as long as the the Wannier functions are invariant under the mirror antisymmetry. 
Second, the TTI is different from the noncompact atomic insulators \cite{Schindler2021}, since the TTI is not an OAI. 
The TTI thus belongs to a new class of non-delta-function insulators with protected gapless edge states.

\textit{Entanglement Spectrum.---} The TTI can be further characterized by the entanglement spectrum (ES) \cite{Li2008,Ryu2006,Turner2010,Fidkowski2010,Hughes2011,Fang2013,Zhu2020,Chang2014}.
The gapless edge states can manifest through the midgap states in the ES \cite{Ryu2006,Turner2010,Fidkowski2010,Hughes2011}. 
This correspondence  works even when the gapless edge states originate from the weak TIs. 
The midgap states make a large contribution to the entanglement entropy. If there exist certain symmetries which protect the midgap states in the ES, the system cannot be adiabatically transformed to a trivial one with zero entanglement entropy. However, the midgap states in the ES can appear even when there is no gapless edge state in the energy spectrum \cite{Hughes2011,Chang2014,Bradlyn2019}, such as in the fragile TIs. 
The spectral flow of the ES, on the other hand, gives an indication to the protected gapless edge states \cite{Hughes2011}. It is represented by the middle bands which continuously flow from the valence bands to the conduction bands in the entanglement energy spectrum. In the case of the strong TIs, the spectral flow necessarily appears due to the presence of protected gapless edge states. \cite{Fidkowski2010,Hughes2011}

\begin{figure}
\begin{center}
{\includegraphics[width=8.6cm]{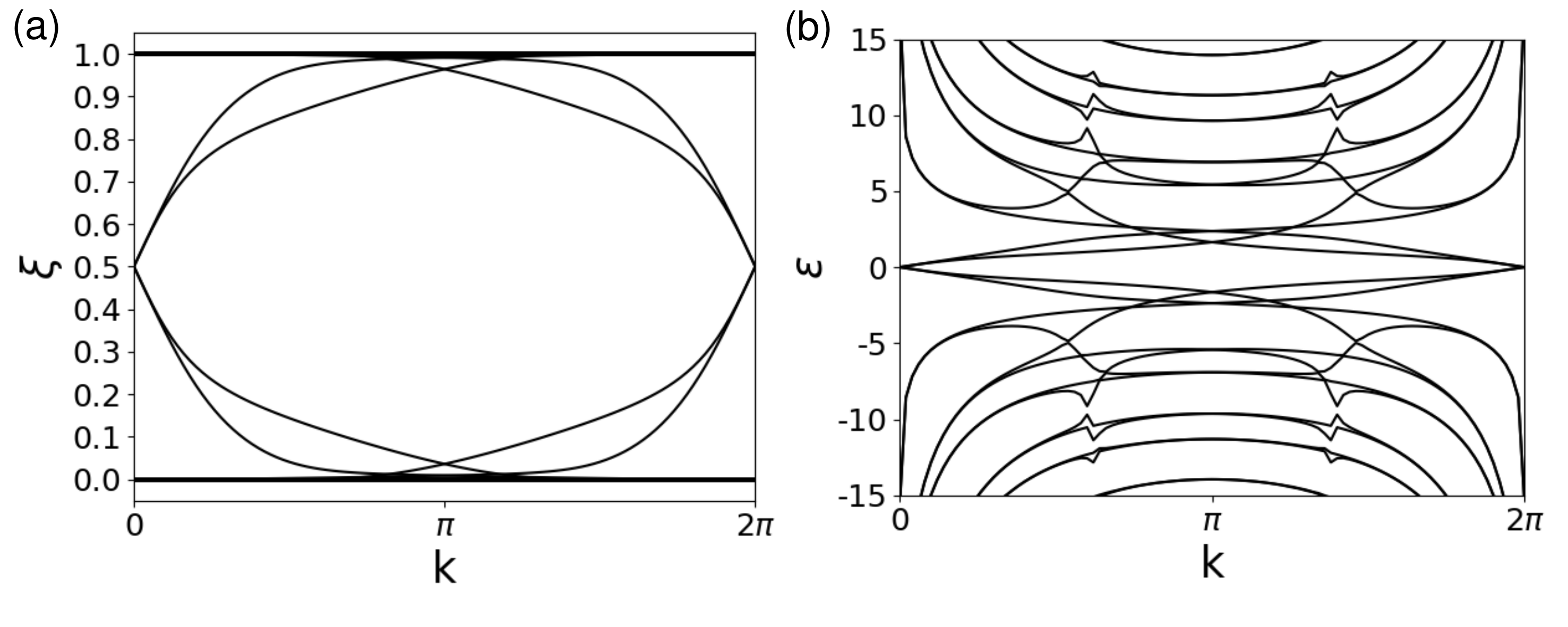}}
\caption[]{Entanglement spectrum of the TTI. Here $m=1.0$ and $B_z=-\lambda_\textrm{SO}=0.7$.
(a) shows the entanglement spectrum.
The entanglement cut is parallel to the $x$-axis. 
(b) shows the entanglement energies. 
}
\label{fig4}
\end{center}
\end{figure} 

Since the ES is equivalent to the spectrum of the flat band Hamiltonian \cite{Fidkowski2010,Hughes2011}, we expect that  nontrivial spectral flows should appear in the TTI.
Figure~\ref{fig4} shows the ES and entanglement energy spectrum of the model (\ref{equation 6}). The ES is calculated from the single-particle correlation function $C_A=\langle c^{\dagger}_ic_j\rangle$, where $i$ and $j$ run over the sites in an equibipartite subregion $A$. The entanglement energy is defined as $\epsilon_n=\log(\xi_n^{-1}-1)/2$, where $\xi_n$ are the $n$-th eigenvalues of $C_A$. 
The spectral flow is clearly observed in the ES and entanglement energy spectrum, as there are four bands flowing from the bottom to the top. 
Note that the TTI serves as a counterexample to an interpolation between the Wilson-loop spectrum and the ES \cite{Lee2015}. Once the $\mathcal{T}$ and $C_2I$ symmetries are broken, the nontrivial entanglement in the model \eqref{equation 6} is protected by the mirror antisymmtry, which does not trigger the the Wannier obstruction or the Wilson-loop winding. Therefore, the nontrivial entanglement does not necessarily imply the Wannier obstruction or Wilson-loop winding since they have different physical origins.
Note that the nontrivial spectral flow can appear in the ES even for the unobstructed atomic insulators with localized Wannier functions at the atoms. 

\textit{Conclusion and discussion.---}In this Letter, we  provide an example of the TTI, where the gapless edge states protected by the mirror antisymmetry exist with trivial topology. These states can be charaterized by the spectral flow in the ES.  The TTI thus belongs to a new class of non-delta-function insulators, yet with stable topology. 

Some remarks regarding the TTI are in order.
First, the TTI is considered relatively topological in  $K$ theory, since it is protected by the mirror antisymmetry which has a $Z$ classification in the Altland-Zirnbauer class $A$ \cite{Geier2018,Trifunovic2019,Shiozaki2014}. 
Therefore, the gapless edge states in the TTI are indeed protected and presumably robust against disorder.
However,  $K$ theory does not  give us the information about the Wannier obstruction. 
It is thus legitimate that the TTI hosts protected gapless edge states while being Wannierizable. 
This is similar to the OAIs, although they come from very different origins.
Interestingly, the topology of the TTI is more trivial, in the sense that the TTI is not an OAI and the Wannier centers are on the atoms. 

Second, our work sheds light on the search for protected gapless edge states in the trivial topology.
Since the TTI can not be captured by the theory of EBRs and Wilson-loop spectrum, it is necessary to develop a systematic approach to characterize  TTIs in  general systems.
As pointed out in the previous literature \cite{Alexandradinata2020,Nelson2021_2}, the multicellularity is necessarily a delicate property if the system is protected by the crystalline symmetry alone. 
In this case, the topological index for the obstruction to constructing delta-like Wannier functions may be hidden in the Wilson-loop spectrum \cite{Nelson2021,Nelson2021_2}. 
On the other hand, the TTI does not manifest Wilson-loop winding, and the chiral symmetry must be included in the study of TTI. A systematic search for  general TTIs is a promising direction for future work. 

\begin{acknowledgments}
The authors thank Yi-Ping Huang, Po-Yao Chang and Chang-Tse Hsieh for useful discussions. Y.C.C. and Y.J.K. were partially supported by the Ministry of Science and Technology (MOST) of Taiwan under grants No. 108-2112-M-002-020-MY3, 110-2112-M-002-034-MY3, 111-2119-M-007-009, and by the National Taiwan University under Grant No. NTU-CC-111L894601.
\end{acknowledgments}

\nocite{*}

\bibliography{citation}

\end{document}


\title{Supplemental Material for ``Protected Gapless Edge States In Trivial Topology"} 
 
\author{Yun-Chung Chen }
\affiliation{Department of Physics and Center for Theoretical Physics, National Taiwan University, Taipei, Taiwan 10617}

\author{Yu-Ping Lin}
\affiliation{Department of Physics, University of Colorado, Boulder, Colorado 80309, USA}
 
\author{Ying-Jer Kao}
\email{yjkao@phys.ntu.edu.tw}
\affiliation{Department of Physics and Center for Theoretical Physics, National Taiwan University, Taipei, Taiwan 10617}
\affiliation{Physics Division, National Center for Theoretical Science, National Taiwan University, Taipei, Taiwan 10617}

\maketitle

\section{Mirror Antisymmetry} \label{sec1}
\vspace{-2ex} 

In this section, we prove several identities about the mirror antisymmetry. Given the low-energy minimal Dirac Hamiltonian,
\begin{eqnarray} \label{equation 1-1}
\mathcal{H}=\bm{k}\cdot\bm{\Gamma}+\lambda M,
\end{eqnarray}
the representation of the mirror antisymmetry squares to 1,
\begin{eqnarray} \label{equation 1-2}
\tilde{\mathcal{M}}_{\bm{z}\times\bm{n}}^2=&&(i\bm{n}\cdot\bm{\Gamma}M)^2 \nonumber\\
=&&-(\bm{n}\cdot\bm{\Gamma})M(\bm{n}\cdot\bm{\Gamma})M \nonumber\\
=&&(\bm{n}\cdot\bm{\Gamma})^2M^2 \nonumber\\
=&&[n_x^2\Gamma_x^2+n_y^2\Gamma_y^2+n_xn_y(\Gamma_x\Gamma_y+\Gamma_y\Gamma_x)] \nonumber\\
=&&1,
\end{eqnarray}
where we have used the fact that the three matrices $\bm{\Gamma}=(\Gamma_x,\Gamma_y)$ and $M$ anticommute with each other and square to identity $I$. 

For any matrix $\Pi$, if $ \{\Pi,\tilde{\mathcal{M}}_{\bm{z}\times\bm{n}}\}=0$, then

\begin{eqnarray} \label{equation 1-3}
P_+(\bm{n})\Pi P_+(\bm{n}) \nonumber=&&\frac{1}{4}(1-\tilde{\mathcal{M}}_{\bm{z}\times\bm{n}})\Pi(1-\tilde{\mathcal{M}}_{\bm{z}\times\bm{n}}) \nonumber\\
=&&\frac{1}{4}(1-\tilde{\mathcal{M}}_{\bm{z}\times\bm{n}}^2) \nonumber\\
=&&0,
\end{eqnarray}
where $P_+(\bm{n})=(1-i\bm{n}\cdot\bm{\Gamma}M)/2=(1-\tilde{\mathcal{M}}_{\bm{z}\times\bm{n}})/2$. Using this property and the fact that $\{\tilde{\mathcal{M}}_{\bm{z}\times\bm{n}},M\}=\{\tilde{\mathcal{M}}_{\bm{z}\times\bm{n}},\bm{n}\cdot\bm{\Gamma}\}=0$, we can obtain the edge Hamiltonian:
\begin{eqnarray} \label{equation 1-4}
&&P_+(\bm{n})(\bm{k}\cdot\bm{\Gamma}+\lambda M)P_+(\bm{n}) \nonumber\\
=&&P_+(\bm{n})k_\parallel(\bm{z}\times\bm{n})\cdot\bm{\Gamma}P_+(\bm{n}) \nonumber\\
\equiv&& k_\parallel\gamma(\bm{n}).
\end{eqnarray}

Finally, we consider the system with the additional chiral symmetry $S$. By definition,

\begin{eqnarray} \label{equation 1-5}
S\mathcal{H}(\bm{k})S^{-1}=-\mathcal{H}(\bm{k}).
\end{eqnarray}

Therefore $S$ must anticommute with the  matrices $\bm{\Gamma}=(\Gamma_x,\Gamma_y)$ and $M$. From this observation, it is evident that
\begin{eqnarray} \label{equation 1-6}
[\tilde{\mathcal{M}}_{\bm{z}\times\bm{n}},S]=0.
\end{eqnarray}
If the system also has the mirror symmetry, the representation of the mirror symmetry that protects the gapless edge states must commute with $S$ since $\tilde{\mathcal{M}}_{\bm{z}\times\bm{n}}=M_{\bm{z}\times\bm{n}}S$. This is consistent with the fact that if the mirror symmetry commutes with the chiral symmetry in class AIII, it has a $Z$ classification and the system hosts protected gapless edge states in the nontrivial phase~\cite{Geier2018,Shiozaki2014}.

\begin{figure}
\begin{center}
\includegraphics[width=8.6cm]{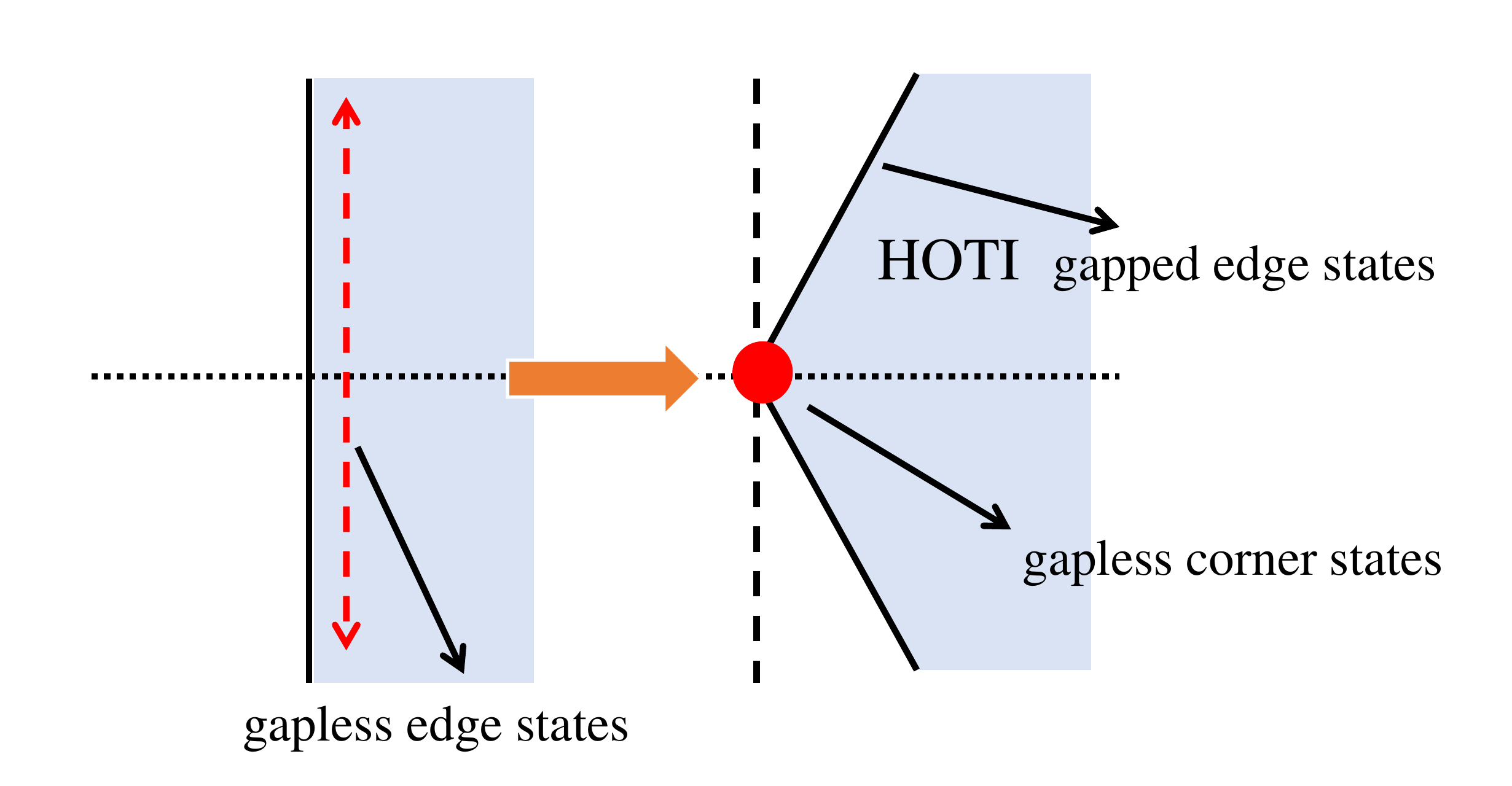}
\caption[]{Relation between the gapless edge states and the gapless corner states in second-order TIs.}
\label{figS1}
\end{center}
\end{figure}

\begin{figure}
\begin{center}
\includegraphics[width=8.6cm]{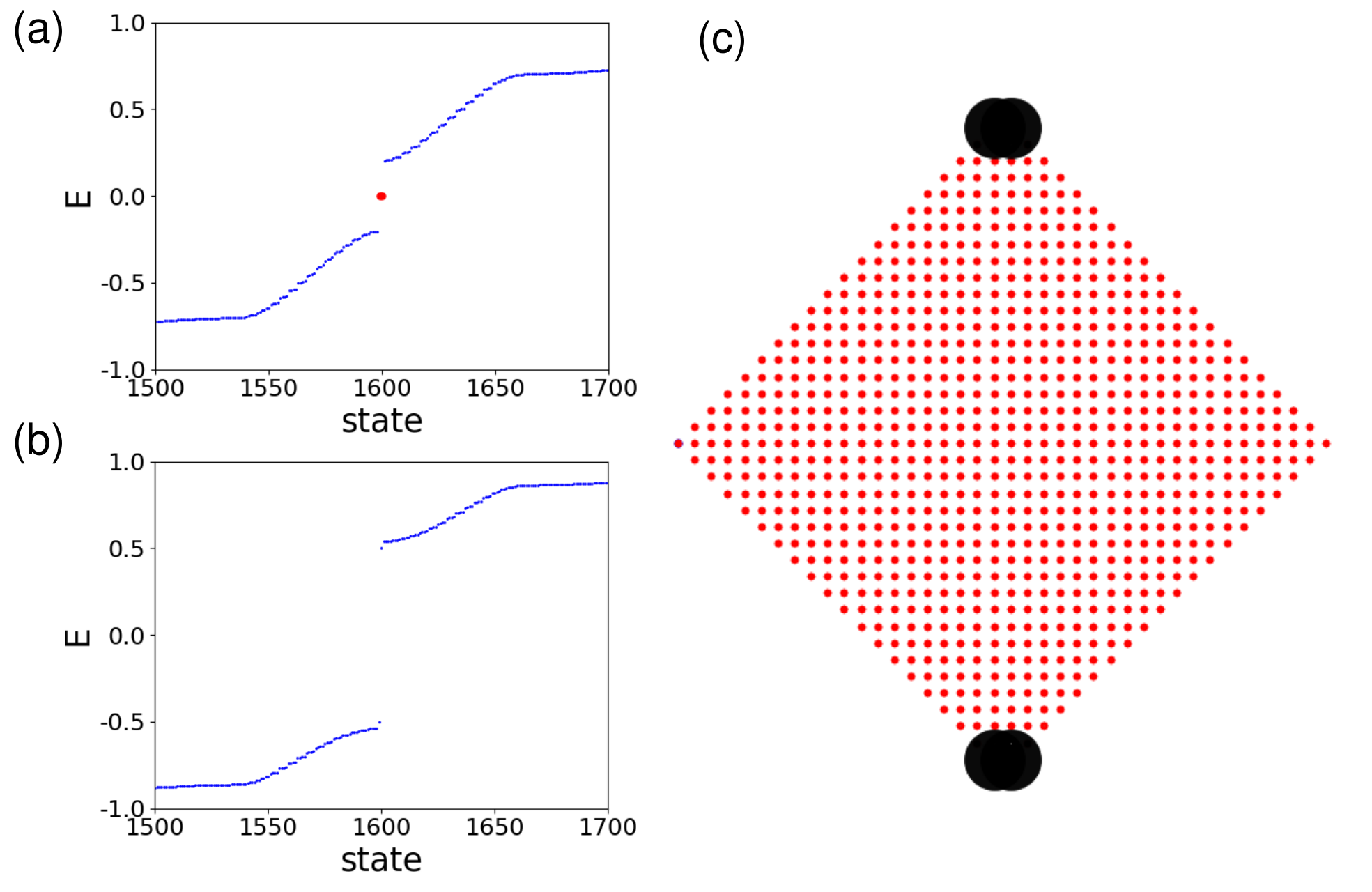}
\caption[]{Second-order topology of the model $H_1$~(\ref{equation 2-4}). (a) shows the nanoflake energy spectrum with $\lambda_1=0$ and $B_{sy}=0.3$. There are two midgap corner states with $E=0$. (c) shows one of the midgap corner states. The corner states are localized at the upper and lower corners of the nanoflake since the gapless edge states only exist at the edges in the $x$-direction. (b) shows the nanoflake energy spectrum with $\lambda_1=-0.5$ and $B_{sy}=0.3$. The midgap corner states completely disappear. Although the mirror symmetry $M_x=is_y$ is still preserved, there is no protecting second-order topology.}
\label{figS2}
\end{center}
\end{figure}

\begin{figure}
\begin{center}
\includegraphics[width=8.6cm]{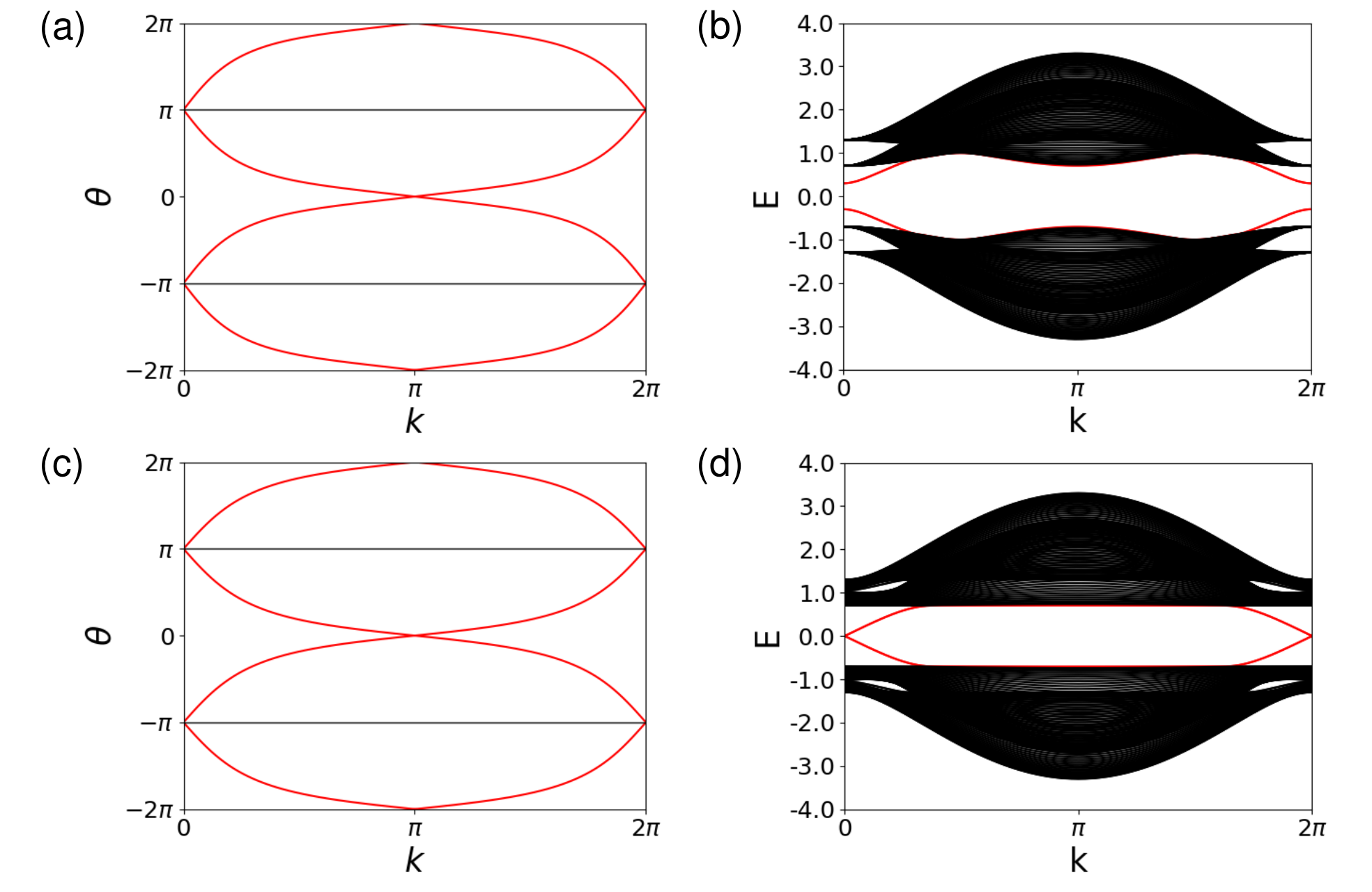}
\caption[]{Wilson-loop spectra of the stable TCI protected by the mirror symmetry $\mathcal{M}_x$ and the chiral symmetries. Here we consider the model $H_1$~(\ref{equation 2-4}) with $\lambda_1=0$ and $B_{sy}=0.3$. (a) and (c) show the Wilson-loop spectra in the $x$ and $y$-direction, respectively. The spectra exhibit relative winding since the system is a stable TCI. (b) and (d) show the nanoribbon energy spectra in the $y$ and $x$-direction, respectively. The gapless edge states only appear in the edges in the $x$-direction since they are protected by the mirror symmetry $\mathcal{M}_x$.}
\label{figS3}
\end{center}
\end{figure}

\section{Higher-order and Fragile Topology} \label{sec2}

In this section, we explicitly construct models and show that contrary to  previous claims~\cite{Ezawa2018, Ren2020, Chen2020}, the gapless edge and corner states are not protected by the mirror~\cite{Ezawa2018,Ren2020} or inversion~\cite{Ezawa2018,Chen2020} symmetry only. Our conclusion is consistent with the classification by both $K$ theory and the layer construction~\cite{Khalaf2018,Geier2018,Trifunovic2019,Hsieh2014}.

The gapless boundary states in 2D can be classified through the order of HOTIs. First-order TIs host protected gapless edge states regardless of the edge terminations. Second-order TIs host midgap states at certain corners. We call these zero-energy corner states as the gapless corner states.

Interestingly, third-order TIs in 2D can also be defined, as in~\cite{Hwang2019}. In general, an $n$-th order TI hosts $n-1$ boundary mass terms regardless of the dimension $d$~\cite{Trifunovic2019,Khalaf2018,Hwang2019}. One can also regard the $n$-th order TIs as systems with protected $(d-n+1)$-dimensional topological surface states~\cite{Benalcazar2017}. These surface states cannot disappear under small perturbations as long as the protecting symmetry is preserved. The third-order TIs in 2D thus host protected topological corner states, which are generically gapped. One can even distinguish the corner states with energy $E>0$ and $E<0$ through the so-called boundary obstructed topological phases~\cite{Khalaf2021}.

\subsection{First-order Topology}

The model in the main text is a time-reversal-invariant $Z_2$ TI:
\begin{eqnarray} \label{equation 2-1}
H(\bm{k})&&=-2\lambda_\textrm{NSO}\sin k_x\sigma_xs_x-2\lambda_\textrm{NSO}\sin k_y\sigma_xs_y \nonumber\\
&&+(m-2t\cos k_x-2t\cos k_y)\sigma_zs_0.
\end{eqnarray}
This model has the same Dirac form as the BHZ model:
\begin{eqnarray} \label{equation 2-2}
H_\textrm{BHZ}(\bm{k})&&=\sin k_x\sigma_xs_z+\sin k_y\sigma_ys_0 \nonumber\\
&&+(2-m-\cos k_x-\cos k_y)\sigma_zs_0.
\end{eqnarray}
Therefore the model~(\ref{equation 2-1}) has a nontrivial $Z_2$ index when $0<m<4t$ and $-4t<m<0$. Notice that the model is also invariant under the $\mathcal{M}_z=C_2I$ internal symmetry. This is a basal mirror symmetry that reflects the 2D plane in the $z$-direction. The associated topological invariant is the mirror Chern number~\cite{Teo2008}, which protects the first-order topology. The Wilson-loop spectra exhibit protected relative winding since each occupied band corresponds to a Chern insulator  with opposite Chern numbers. The Wilson-loop and edge energy spectra correspondence is thus satisfied. In this case, the system is protected by an internal symmetry and therefore also belongs to the strong TIs. 

\subsection{Second-order Topology}

It is pointed out that there is an edge-corner correspondence in second-order topology, see Fig.~\ref{figS1}~\cite{Geier2018}. One can typically find a high-symmetry edge with gapless edge states in second-order TIs. This is again related to the allowed one boundary mass term due to the protecting symmetry. 

The model (\ref{equation 2-1}) becomes a stable mirror-symmetry protected TCI if we preserve both the chiral symmetry $S=-i\sigma_y$ and the mirror symmetries. Since a mirror antisymmetry is a combination of the chiral and mirror symmetries, one can obtain the representations of the mirror symmetries that protect the gapless edge states,
\begin{eqnarray} \label{equation 2-3}
\mathcal{M}_x=-is_y, \nonumber\\
\mathcal{M}_y=-is_x.
\end{eqnarray}
From this observation, it is clear that if the system has the chiral symmetry, the mirror symmetry is equivalent to the mirror antisymmetry. 

However, as mentioned in the main text, the mirror symmetry alone cannot protect the gapess edge states and the second-order topology. As a simple demonstration, we break the chiral symmetry $S=-i\sigma_y$ but preserve the mirror symmetry $\mathcal{M}_x=-is_y$ in the $x$-direction. This can be done by adding the terms,
\begin{eqnarray} \label{equation 2-4}
H_{1}=&&H+\lambda_{1}\sum_{i,a,b}\zeta_{ab}c^{\dagger}_{i,a}c_{i,b}\nonumber\\
&&+B_{sy}\sum_{i,a}c^{\dagger}_{i,a}\xi_{a}s_yc_{i,a},
\end{eqnarray}
where $\zeta_{ab}=\sigma^y_{ab}$ and $\xi_a=\pm1$ for orbitals $a=A,B$. The second term is the in-plane staggered-ordered field $B_{sy}$ in the $y$-direction. This term serves as a boundary mass term in the $y$-direction such that the gapless corner states can appear. One can easily check that this term breaks the mirror antisymmetry $\tilde{\mathcal{M}}_y$. When $\lambda_1=0$, this system is a stable second-order TCI protected by the mirror symmetry $\mathcal{M}_x$ and the chiral symmetries $S$. However, when $\lambda_1\neq 0$, the first term breaks the chiral symmetry and serves as a bulk mass term since it anticommutes with the three matrices $\bm{\Gamma}=(\Gamma_x,\Gamma_y)$ and $M$. Therefore, the midgap states disappear and the edges are fully gapped, see Fig.~\ref{figS2}.

\begin{figure}
\begin{center}
\includegraphics[width=8.6cm]{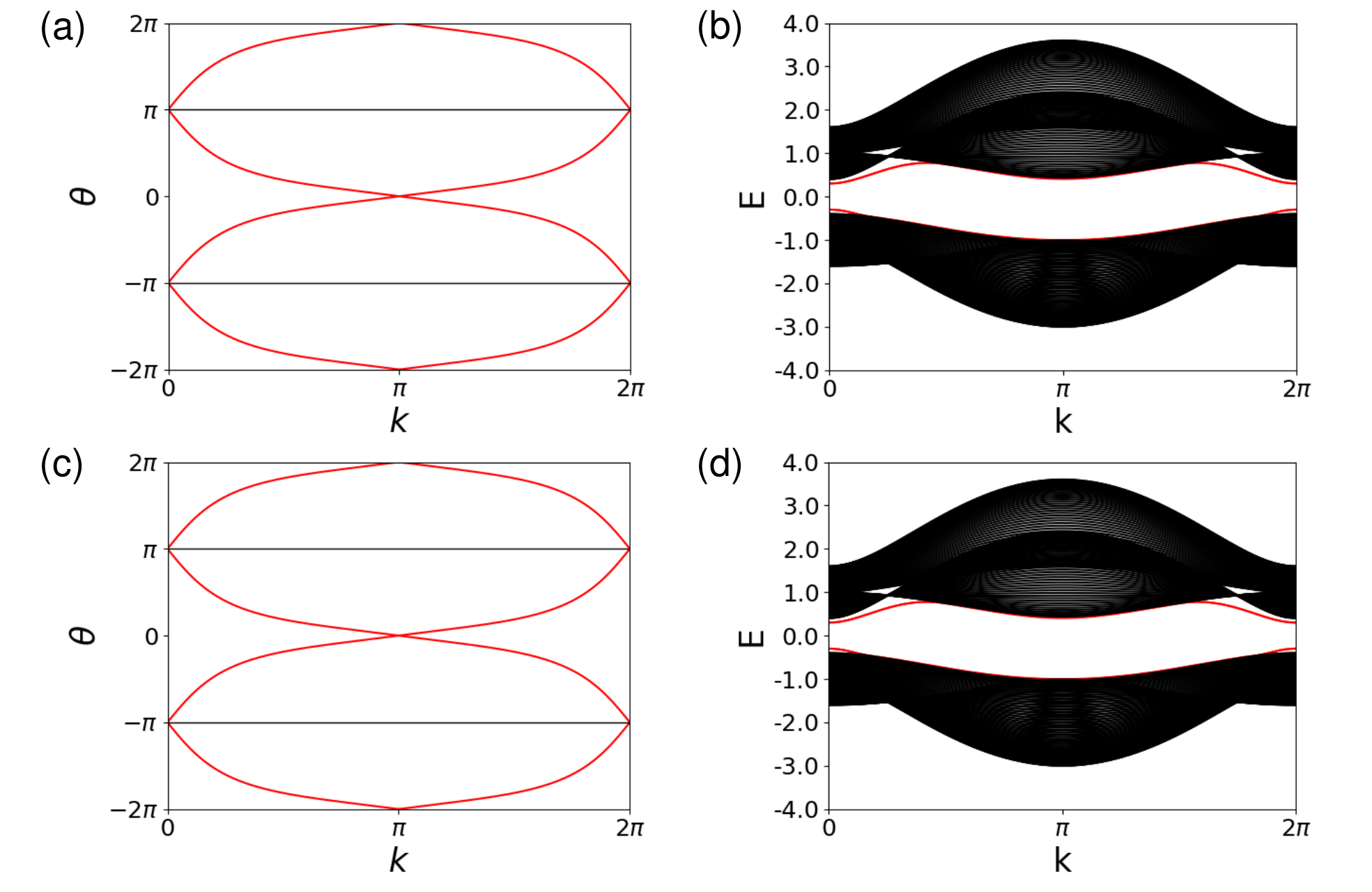}
\caption[]{Third-order and fragile topology of model $H_2$~(\ref{equation 2-5}) with $B_{sy}=B_y=0.3$. (a) and (c) show the Wilson-loop spectra in the $x$ and $y$-direction, respectively. The spectra exhibit relative winding since the system is a fragile TI. (b) and (d) show the nanoribbon energy spectra in the $y$ and $x$-direction, respectively. Both the edge states become gapped due to the two different boundary mass terms.}
\label{figS4}
\end{center}
\end{figure}

Now we examine the Wilson-loop spectra of the model~(\ref{equation 2-4}) with $\lambda_1=0$. Since it is a stable TCI protected by the mirror and chiral symmetries, the relative winding of the Wilson-loop spectrum is protected, see Fig.~\ref{figS3}~\cite{Neupert2018}. Notice that the edge states in the $y$-direction are gapped, therefore the Wilson-loop and edge energy spectra correspondence is broken in the $y$-direction. One may argue that the correspondence fails since the mirror symmetry $\mathcal{M}_y$ which protects the correspondence is broken~\cite{Alexandradinata2016}. We note that the system is still invariant under another mirror symmetry $\mathcal{M}'_y=-\sigma_zis_y$. This mirror symmetry leaves the Wilson-loop and edge energy spectra invariant in the $y$-direction. Therefore, even if the transformation is invariant under the symmetry, the Wilson-loop and edge energy spectra correspondence can still be broken.

\subsection{Third-order and Fragile Topology}

Third-order topology can be produced by adding two boundary mass terms that preserve the inversion ($\mathcal{I}$) symmetry:
\begin{eqnarray} \label{equation 2-5}
H_{2}=&&H+B_{sy}\sum_{i,a}\xi_{a}c^{\dagger}_{i,a}s_yc_{i,a}\nonumber\\
&&+B_y\sum_{i,a}c^{\dagger}_{i,a}s_yc_{i,a}.
\end{eqnarray}
Notice that these two terms break the mirror antisymmetries $\tilde{\mathcal{M}}_x$ and $\tilde{\mathcal{M}}_y$. As mentioned in the main text, the system has a fragile topology protected by the inversion symmetry. In this case, the gapless edge states disappear due to the two boundary mass terms, see Fig~\ref{figS4}. The Wilson-loop spectra exhibit relative winding due to the Wannier obstruction \cite{Hwang2019}. Therefore, the Wilson-loop and edge energy spectra correspondence breaks completely in the fragile topology.

\section{Wannier Function} \label{sec3}

In this section, we explicitly construct the $C_4$-symmetric exponentially-localized Wannier functions of the TTI model. In addition, the Wannier functions are invariant under the mirror antisymmetries: they form the mirror pairs between the conduction and valence bands.

\subsection{Pairing condition}
Here we consider the Wannier functions in the presence of the point-group (PG) antisymmetry, which is a combination of the order-two point group and chiral symmetries. Assume the system is also invariant under the space group (SG) symmetry $G$, and the bands are always half-filling. 

First we consider the point group of the antisymmetry ($\tilde{G}$) is a subgroup of the space group $G$. In this case, it is equivalent to adding an additional chiral symmetry to the system. The valence band Wannier functions should satisfy the chiral pairing condition:
\begin{eqnarray} \label{equation 3-1}
\langle W^v_{\bm{R},m}|A_S|W^v_{\bm{0},n}\rangle=0,
\end{eqnarray}
for any valence band index $n,m$ and positions $\bm{R}$, where $A_S$ is the chiral operator in real space. Consider the Bloch states,
\begin{eqnarray}
e^{-i\bm{k}\cdot\bm{r}}|u^v_n(\bm{k})\rangle=\sum_{\bm{R}}e^{i\bm{k}\cdot\bm{R}}|W^v_{\bm{R},n}\rangle,
\end{eqnarray}
then the matrix elements
\begin{eqnarray}
\langle u^v_m(\bm{k})|S|u^v_n(\bm{k})\rangle=0
\end{eqnarray}
become zero for any valence band index $n,m$ and lattice momentum $\bm{k}$ due to the condition~(\ref{equation 3-1}), where $S$ is the representation of the chiral operator. Therefore, the states
\begin{eqnarray}
|u^c_n(\bm{k})\rangle=S|u^v_n(\bm{k})\rangle
\end{eqnarray}
must belong to the conduction bands and form the basis of the conduction band wavefunctions. It is shown in the main text that this pairing condition is enough to characterize the gapped Wannierlizable Hamiltonians with the chiral symmetry.

If the Hamiltonian has the chiral symmetry and one of the conduction and valence bands are band representations (BRs), then the two quasiband representations (qBRs) must be BRs. In other words, it is possible to find the exponentially-localized  symmetric Wannier functions that also obey the chiral symmetry. Without loss of generality, we consider the valence band Wannier functions at $\bm{R}=\bm{0}$:
\begin{eqnarray} \label{equation 3-2}
&&|W_{\bm{0},n\in v}\rangle \nonumber\\
=&&\frac{V}{(2\pi)^d}\int_{BZ}d^d\bm{k}e^{-i\bm{k}\cdot\bm{r}}\sum_mU_{nm}(\bm{k})|u_{m\in v}(\bm{k})\rangle,
\end{eqnarray}
where $d$ is the dimension of the system, and $V$ is the volume of the unit cell. $|u_{i,m}(\bm{k})\rangle$ is the periodic part of the valence band Bloch states, and $U_{nm}(\bm{k})$ is a $U(N)$ gauge transformation such that the Wannier functions are exponentially-localized and obey the space group symmetry $G$. Now we act the chiral operator on $|W_{\bm{0},n\in v}\rangle$:
\begin{eqnarray} \label{equation 3-3}
&&A_S|W_{\bm{0},n\in v}\rangle \nonumber\\
=&&\frac{V}{(2\pi)^d}\int_{BZ}d^d\bm{k}e^{-i\bm{k}\cdot\bm{r}}\sum_mU_{nm}(\bm{k})S|u_{m\in v}(\bm{k})\rangle\nonumber\\
=&&\frac{V}{(2\pi)^d}\int_{BZ}d^d\bm{k}e^{-i\bm{k}\cdot\bm{r}}\sum_mU_{nm}(\bm{k})|u_{m\in c}(\bm{k})\rangle\nonumber\\
=&&|W_{\bm{0},n\in c}\rangle.
\end{eqnarray}
Notice that the overlap matrix is also mapped to the conduction bands:
\begin{eqnarray} \label{equation 3-4}
\delta_{kl}=&&\sum_{nm}\langle u_{n\in v}(\bm{k})|U^{\dagger}_{nk}(\bm{k})U_{lm}(\bm{k})|u_{m\in v}(\bm{k})\rangle \nonumber\\
=&&\sum_{nm}\langle u_{n\in v}(\bm{k}) |U^{\dagger}_{nk}(\bm{k})S^{\dagger}SU_{lm}(\bm{k})|u_{m\in v}(\bm{k})\rangle\nonumber\\
=&&\sum_{nm}\langle u_{n\in c}(\bm{k})|U^{\dagger}_{nk}(\bm{k})U_{lm}(\bm{k})|u_{m\in c}(\bm{k})\rangle.
\end{eqnarray}
Therefore, $U_{lm}$ is also the proper gauge transformation for constructing the conduction band Wannier functions. The conduction and valence bands are thus both Wannierlizable and respect the chiral pairing condition.

In the case of the stable TCI discussed in Sec.~\ref{sec2},  valence qBRs are not Wannier-representable under the mirror symmetry as long as the Hamiltonian obeys the chiral symmetry. This is due to the fact that the system has protected Wilson-loop windings~\cite{Neupert2018}. This serves as an obstruction when constructing the mirror-symmetric Wannier functions. Note that this obstruction cannot be detected by the irreducible representations of the little groups, or the symmetry indicators. This is because one can preserve the mirror symmetry but break the chiral symmetry to adiabatically deform the system into a trivial atomic insulator. This is a universal feature of the chiral-symmetry protected TIs or TCIs: they can only be detected by the windings of the Wilson loop.

\begin{figure}
\begin{center}
\includegraphics[width=8.6cm]{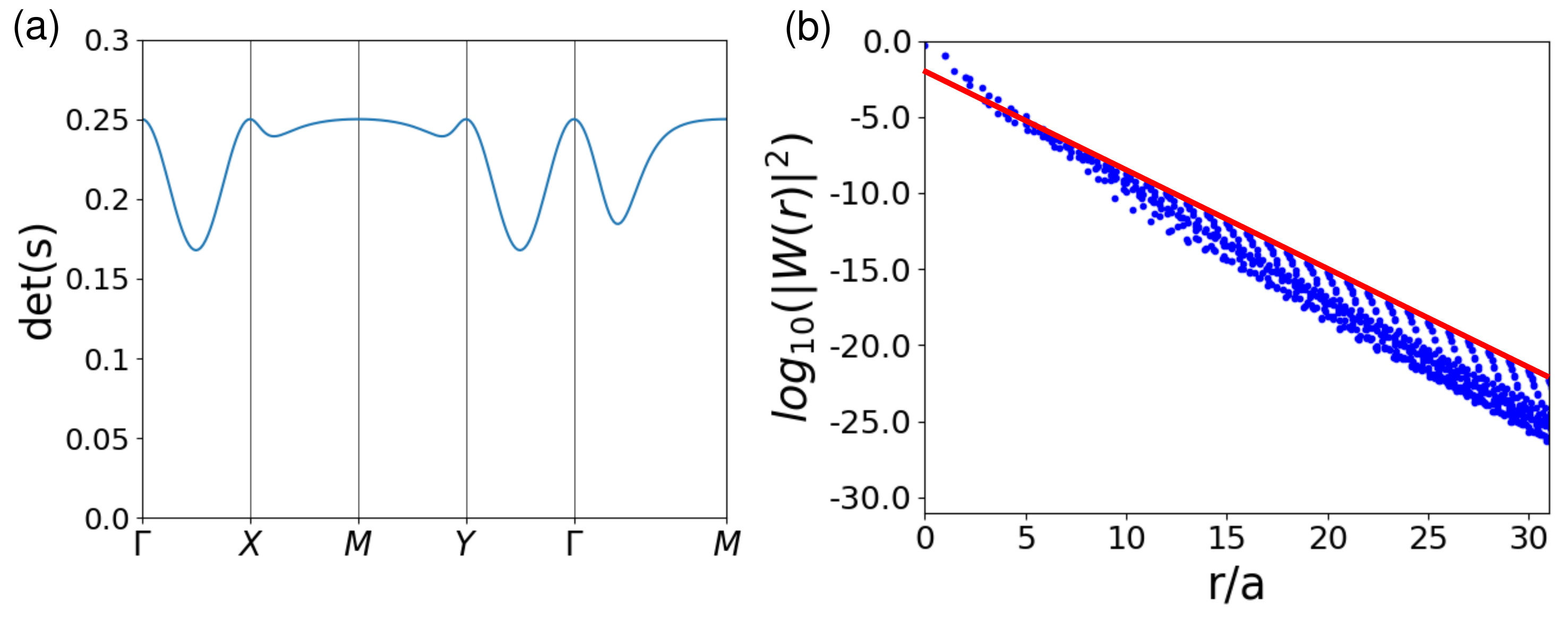}
\caption[]{(a) $\det[s_{nm}(\bm{k})]$ of the TTI. Here we set the parameters $m=1.0$ and $B_z=-\lambda_\textrm{SO}=0.7$. Since $\det[s_{nm}(\bm{k})]>0$ for all $\bm{k}$ in the Brillouin zone, there is no obstruction to constructing the exponentially-localized Wannier functions. (b) The exponential localization of the fist valence band Wannier function $W^v_1(r)$. The probability of the Wannier wavefunction exponentially approaches to zero when $r$ is large. The other Wannier functions also show the same decaying behavior.}
\label{figS5}
\end{center}
\end{figure}

\begin{figure}
\begin{center}
{\includegraphics[width=8.6cm]{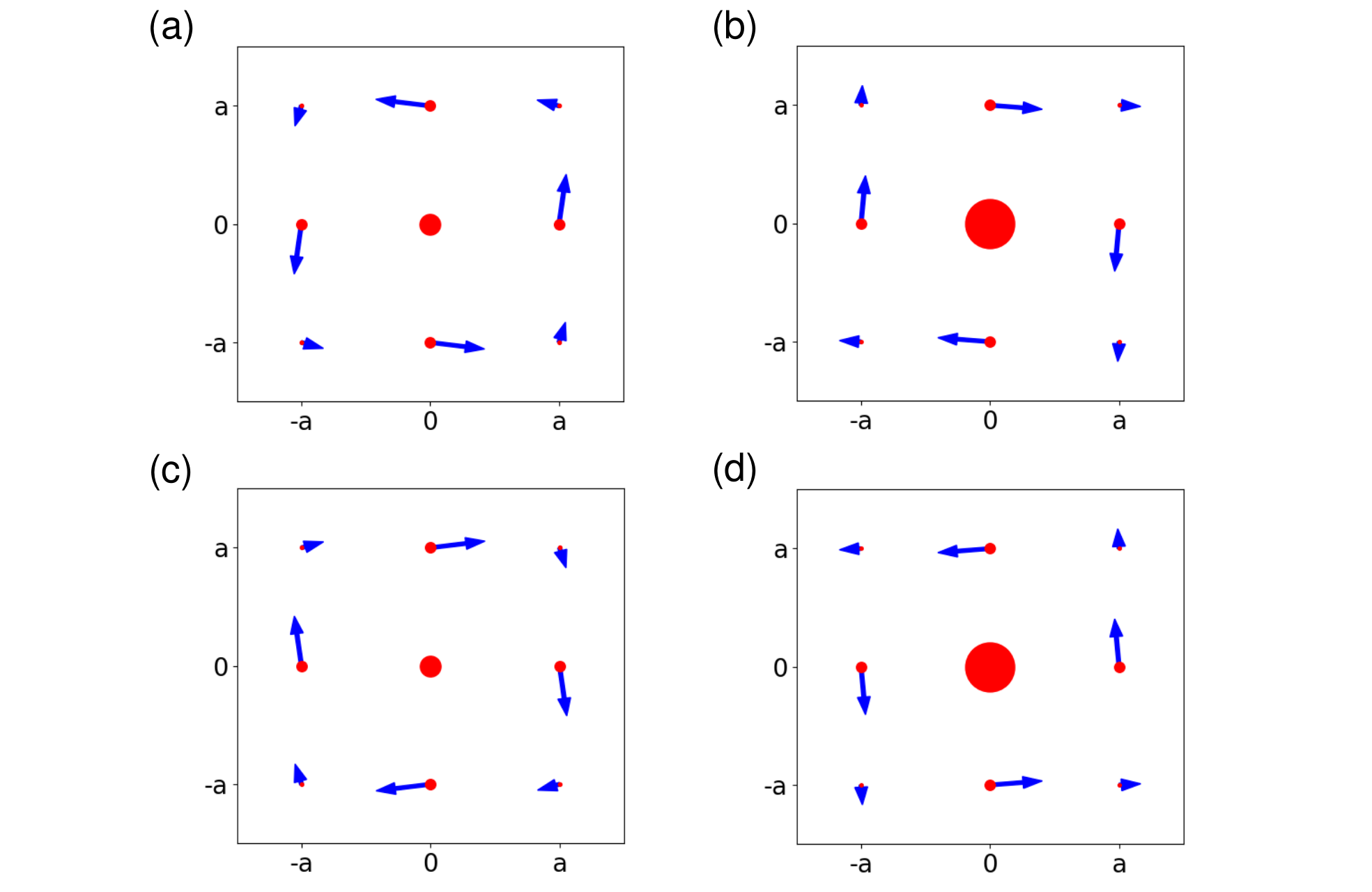}}
\caption[]{The second Wannier functions of our TTI model. They are all localized at the Wyckoff position $a$ and respect the $C_4$ symmetry. Each plot corresponds to a single Wannier function. Here we only plot the neighboring nine atoms, although the results are more extended. The sizes of the red spots are proportional to the probability of the Wannier functions. The blue arrows represent the magnitude and direction of the spins in the $x-y$ plane. (a) and (b) illustrate one of the valence band Wannier functions $|W_{\bm{0},2}^v\rangle$ projected to orbital $A$ and $B$, respectively. (c) and (d) is the corresponding conduction band Wannier function $|W_{\bm{0},2}^c\rangle$ projected to orbital $B$ and $A$, respectively. It is clear that the Wannier functions of (a) and (c), (b) and (d) form the mirror pairs. Therefore, they also satisfy the mirror antisymmetry.}
\label{figS6}
\end{center}
\end{figure}

For the case that the point group of the antisymmetry ($\tilde{G}$) is not a subgroup of the space group $G$, the antisymmetric pairing condition is:
\begin{eqnarray} \label{equation 3-5}
\langle W^v_{\bm{R},m}|A_{\tilde{G}}|W^v_{\bm{0},n}\rangle=0,
\end{eqnarray}
for any valence band index $n,m$ and positions $\bm{R}$, where $A_{\tilde{G}}$ is the $\tilde{G}$-antisymmetry operator in real space. The group $\tilde{G}=\{I,g\}\cong Z_2$. 

One can follow the same idea to prove that if the Hamiltonian has the $\tilde{G}$-antisymmetry and one of the conduction or valence bands are BRs, then the two qBRs must be BRs. The only difference is that $g_{\tilde{G}}|u_{m\in v}(\bm{k})\rangle=|u_{m\in c}(R\bm{\bm{k}})\rangle$, where $g_{\tilde{G}}$ is the representation of $g=\{R|\bm{0}\}$. The paired Wannier functions are thus functions of $R^T\bm{r}=R^{-1}\bm{r}=R\bm{r}$. Therefore, the valence and conduction band Wannier functions also satisfy the $\tilde{G}$-antisymmetry.

\subsection{Trivialized Topological Insulator}

Contrary to the stable TCI discussed in Sec.~\ref{sec2}, the TTI encounters no Wannier obstruction since the Wilson loop has no protected winding. It is thus possible to construct the Wannier functions that respect the space group symmetry $P4$ and obey the antisymmetric pairing condition~(\ref{equation 3-5}). 

We use the standard projection method to construct the Wannier functions~\cite{Marzari2012}. Start with two trial wavefunctions $|\tau_n(\bm{k})\rangle$ which are the Fourier transforms of two $C_4$-symmetric atomic Wannier functions at Wyckoff position $a$, we project them into the occupied space,
\begin{eqnarray} \label{equation 3-6}
|\chi_n(\bm{k})\rangle=P(\bm{k})|\tau_n(\bm{k})\rangle,
\end{eqnarray}
where $P(\bm{k})=\sum_n|u_n(\bm{k})\rangle\langle u_n(\bm{k})|$ is the projector onto the valence bands. The Löwdin orthogonalization procedure is used,
\begin{eqnarray} \label{equation 3-7}
s_{nm}(\bm{k})=\langle \chi_n(\bm{k})|\chi_m(\bm{k})\rangle,
\end{eqnarray}
and if this matrix has nonvanishing determinant for all $\bm{k}$, we can define a set of orthonormal basis,
\begin{eqnarray} \label{equation 3-8}
|u_n(\bm{k})\rangle=\sum_m[s(\bm{k})]^{-1/2}_{nm}|\chi_m(\bm{k})\rangle,
\end{eqnarray}
and use this set to construct the Wannier functions according to Eq.~(\ref{equation 3-2}). The obstruction may occur when the determinant $\det[s_{nm}(\bm{k})]=0$ for some $\bm{k}$, and the orthogonalization procedure is failed. In the case of the Chern insulator, there is no choice for the trial wavefunctions $|\tau_n(\bm{k})\rangle$ satisfying $\det [s_{nm}(\bm{k})]>0$ for all $\bm{k}$, therefore there is an obstruction to constructing the exponentially-localized Wannier functions. However, in the model of our TTI, no obstruction is found since $\det [s_{nm}(\bm{k})]>0$ for all $\bm{k}$, see Fig.~\ref{figS5}. In the main text, we have already plotted the Wannier functions $|W^v_{\bm{0},1}\rangle$ and $|W^c_{\bm{0},1}\rangle$. Here we plot the Wannier functions $|W^v_{\bm{0},2}\rangle$ and $|W^c_{\bm{0},2}\rangle$ in Fig.~\ref{figS6}. They also obey the $C_4$ symmetry and the mirror antisymmetry.

\nocite{*}

\bibliography{citation}